\documentclass{article}

\usepackage[preprint]{neurips_2026}

\usepackage[utf8]{inputenc}
\usepackage[T1]{fontenc}
\usepackage{hyperref}
\hypersetup{hidelinks}
\usepackage{url}
\usepackage{booktabs}
\usepackage{amsfonts}
\usepackage{amsmath}
\usepackage{nicefrac}
\usepackage{microtype}
\usepackage{xcolor}
\usepackage{graphicx}
\usepackage{subcaption}
\usepackage{wrapfig}
\usepackage[skip=3pt]{caption}
\usepackage{placeins}
\usepackage{float}
\usepackage[most]{tcolorbox}
\definecolor{promptbg}{HTML}{FFF6E0}

\newtcolorbox{promptbox}[1][]{%
  enhanced, breakable,
  colback=promptbg, colframe=black, boxrule=0.5pt,
  arc=3pt, left=8pt, right=8pt, top=6pt, bottom=6pt,
  fonttitle=\bfseries\small,
  coltitle=black,
  attach boxed title to top left={xshift=8pt, yshift=-7pt},
  boxed title style={colback=white, colframe=black, arc=2pt, boxrule=0.4pt},
  #1
}

\graphicspath{{figures/}}

\title{Reward and Behavioral Measures of LLM Coordination in Dining Philosophers}

\author{
Najmul Hasan \quad Prashanth BusiReddyGari\thanks{
Corresponding author: \texttt{prashanth.busireddygari@uncp.edu}.\\ 
Code is available at: \url{https://github.com/najmulhasan-code/dpmarl}
} \\
Department of Mathematics and Computer Science \\
University of North Carolina at Pembroke \\
Pembroke, NC, USA
}

\begin{document}

\maketitle

\begin{abstract}
Does high reward indicate productive coordination among language-model agents sharing resources? We examine this question in dining philosophers, where instances of the same model must acquire shared forks to eat. We evaluate seven model conditions across 630 episodes at five, ten, and fifteen philosophers, measuring scalar reward, deadlock, and meal throughput. At five philosophers, Mistral-Small 24B receives mean reward 0.827 while producing no meals. The reward decomposition explains this mismatch: avoiding deadlock earns a bonus without requiring meal production. To assess a policy trained with this reward, we compare the final checkpoint from a single Qwen3-14B group relative policy optimization (GRPO) training run with its base model. At five philosophers, the trained model has lower throughput (0.008 versus 0.020) and a higher deadlock rate in point estimates. Differences in mean episode reward are not statistically significant at any evaluated philosopher count. These findings show that the evaluated reward can assign high scores without task progress and support reporting behavioral measures alongside reward when comparing models and evaluating fine-tuned policies.
\end{abstract}

\begin{figure}[h]
\centering
\includegraphics[width=\linewidth]{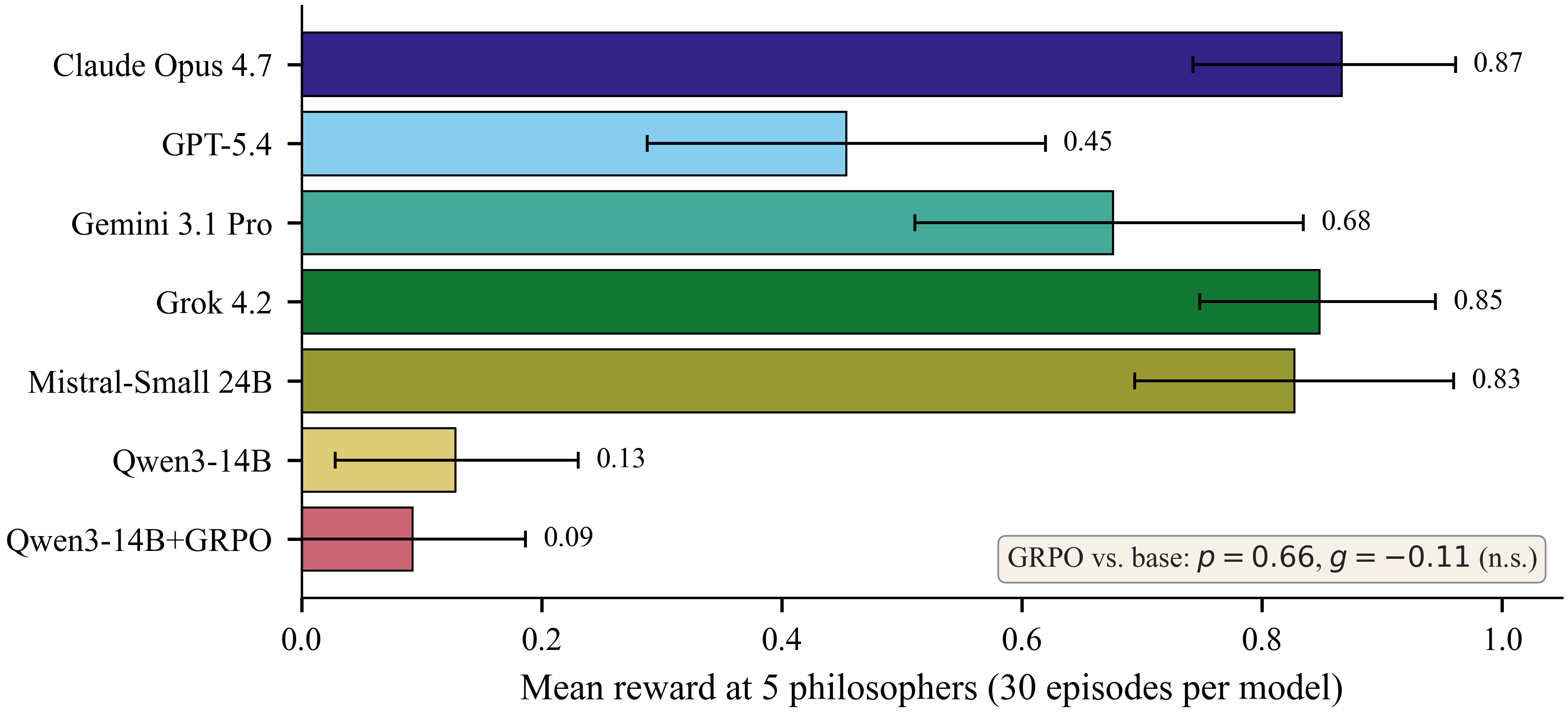}
\caption{Mean reward at five philosophers. Mistral-Small 24B receives 0.83 with zero throughput (Table~\ref{tab:main-results}), so high reward need not indicate productive coordination. The saved Qwen3-14B+GRPO versus base reward comparison gives $p=0.66$ in the reported Welch test; this does not establish equivalence.}
\label{fig:teaser}
\end{figure}

\section{Introduction}
\label{sec:intro}

Evaluating agents that share resources requires distinguishing productive resource use from the absence of deadlock. In our dining-philosophers environment, an episode can end without triggering the deadlock condition even though no philosopher has eaten. A reward that combines deadlock avoidance and meal production into a single score can conceal that distinction. We use this setting to examine what scalar reward and its behavioral components reveal about tool-using language models.

The environment assigns all philosophers to instances of the same policy model. Each philosopher must acquire both adjacent forks to eat, and the environment schedules one turn per philosopher in a shuffled order each round. This makes resource ownership and meal production explicit while retaining the model's responsibility for choosing tool calls. We measure deadlock, throughput, meal-count variance, and the fraction of think actions alongside the combined reward.

\begin{samepage}
Our main evaluation covers seven model conditions and 630 episodes: four closed-source models, Mistral-Small 24B, and Qwen3-14B~\cite{yang2025qwen3} before and after group relative policy optimization (GRPO) fine-tuning. Each condition is evaluated at three philosopher counts. A separate experiment evaluates Qwen3-8B, its trained variant, and DeepSeek-R1-Distill-Qwen-7B across philosopher counts and episode lengths. Our primary question is whether scalar reward reflects productive behavior in these evaluations. The training experiment uses the same reward, making its interpretation relevant both to cross-model comparisons and to assessing the saved policy against its base.
\par
\end{samepage}

\textbf{Reward and task progress.}
At five philosophers, Mistral-Small 24B receives mean reward 0.827 with zero throughput (Table~\ref{tab:main-results}). The four closed-source models have positive throughput, ranging from 0.057 to 0.105. Thus, Mistral's high reward cannot be read as evidence of comparable meal production. Figure~\ref{fig:teaser} presents the reward comparison; the component measures explain why its ordering is insufficient to rank productive coordination. The reward definition and the serving-dependent DeepSeek-R1 result provide further examples of reward without meals.

\textbf{The saved-policy comparison.}
We examine the saved Qwen3-14B+GRPO policy using reward together with throughput and deadlock. At five philosophers, it has lower throughput and higher deadlock than its base in point estimates. The reported reward tests detect no significant difference at any evaluated philosopher count. This is a comparison of the saved policy from one training run with its base, not a test of GRPO's effectiveness across training runs or reward designs.

The main contribution is an empirical demonstration that this reward can assign high scores without meal production, together with an explanation from its component terms. The saved-policy comparison provides a second result under the same measurement framework: reward tests and behavioral point estimates describe different aspects of the evaluated policy. Sections~\ref{sec:task}--\ref{sec:experiments} describe the task and experimental setup; Sections~\ref{sec:results}--\ref{sec:limitations} present the results, their interpretation, and the limits of the evidence.

\section{Related work}
\label{sec:related}

\textbf{Tool use and agent evaluation.}
ReAct~\cite{yao2023react} combines reasoning with actions, while Toolformer~\cite{schick2023toolformer} studies learning to use tools. AgentBench~\cite{liu2024agentbench}, WebArena~\cite{zhou2024webarena}, SWE-bench~\cite{jimenez2024swebench}, GAIA~\cite{mialon2024gaia}, and $\tau$-bench~\cite{yao2025taubench} evaluate agents across interactive tasks, including web use, software engineering, and tool--user interaction. HELM~\cite{liang2023helm} provides a broader framework for evaluating language models across scenarios and metrics. Our evaluation focuses on a specific interaction: several instances of one model acquire and release shared resources. Reporting meals and deadlock separately makes it possible to examine whether the scalar score reflects progress in that interaction.

\textbf{Multiple agents and coordination.}
Multi-agent debate~\cite{du2024multiagent}, CAMEL~\cite{li2023camel}, AutoGen~\cite{wu2024autogen}, and MetaGPT~\cite{hong2024metagpt} examine or support collaboration among language-model agents. Generative Agents~\cite{park2023generative} studies simulated social behavior. Coordination also appears in settings such as Hanabi~\cite{bard2020hanabi}, Overcooked~\cite{carroll2019overcooked}, and Diplomacy~\cite{meta2022cicero}. Dining philosophers~\cite{dijkstra1971hierarchical} provides a setting centered on shared-resource acquisition: an agent's ability to eat depends on fork ownership, including the actions of its neighbors.

\textbf{Reinforcement learning and policy adaptation.}
Reinforcement learning from preferences~\cite{christiano2017rlhf} and instruction tuning with human feedback~\cite{ouyang2022instructgpt} use feedback to train policies. PPO~\cite{schulman2017ppo} and GRPO~\cite{shao2024deepseekmath} provide approaches to policy optimization; GRPO uses relative rewards within rollout groups and is also used in DeepSeek-R1~\cite{deepseekai2025r1}. Our experiment applies GRPO to an environment-computed reward for dining philosophers. It evaluates the saved policy against its base, without comparing optimization algorithms or varying the reward design.

\textbf{Fine-tuning and serving.}
We use low-rank adaptation (LoRA)~\cite{hu2022lora} for the Qwen3 training runs and vLLM~\cite{kwon2023vllm} to serve the open-weight policies. These components matter to the interpretation of the experiment: the trained policy includes a LoRA adapter, and the environment depends on outputs being parsed as tool calls. The DeepSeek-R1-Distill-Qwen-7B result also depends on the reported tool-call parser.

\section{The dining philosophers task and reward}
\label{sec:task}

\begin{wrapfigure}{R}{0.34\linewidth}
\centering
\vspace{-18pt}
\includegraphics[width=\linewidth]{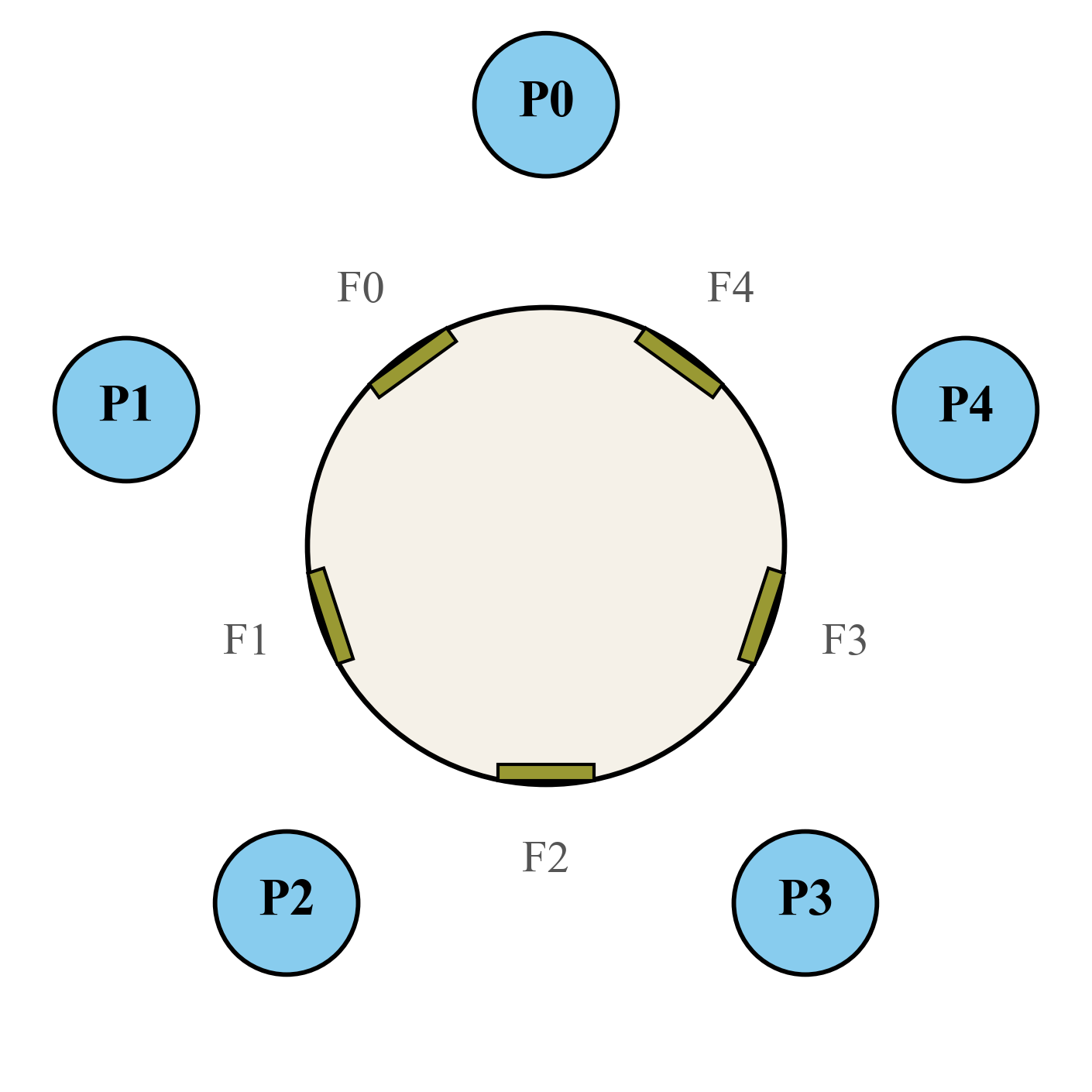}
\caption{Five philosophers seated around a table share five forks; a philosopher can eat only when holding both adjacent forks.}
\label{fig:task}
\vspace{-12pt}
\end{wrapfigure}

We adopt the dining philosophers problem~\cite{dijkstra1971hierarchical} as a coordination benchmark for tool-using LLM agents. $N$ philosophers sit around a circular table. Between each pair of adjacent philosophers lies a single fork, for $N$ forks total. Each philosopher must hold both adjacent forks simultaneously to eat. Episodes proceed in rounds; each round schedules one turn per philosopher in a randomly shuffled order. The episode ends when all rounds complete or when the environment's reported deadlock stopping condition is detected: every philosopher holds exactly one fork and is waiting for the other. We use deadlock to refer to this stopping condition. Figure~\ref{fig:task} illustrates the setup.

Each philosopher is exposed as an LLM agent with six tools: pick-up-left, pick-up-right, put-down-left, put-down-right, eat, and think. The eat action is admissible only when both adjacent forks are held; put-down actions are admissible only for forks the philosopher currently holds. The agent receives a static system prompt that fixes its role and adjacent fork layout, and a per-turn human message that reports the current state of its two adjacent forks and its meals so far; it is instructed to return a single tool call per turn. The exact prompts and tool descriptions are reproduced verbatim in Appendix~\ref{app:reproducibility}.

We score each episode with a four-component scalar reward,
\begin{equation}
r = \alpha \cdot r_{\text{nd}} + \beta \cdot r_{\text{thru}} - \gamma \cdot r_{\text{fair}} - \delta \cdot r_{\text{idle}},
\label{eq:reward}
\end{equation}
where $r_{\text{nd}}\in\{0,1\}$ indicates the absence of deadlock at termination. Let $T$ be the configured number of rounds and $m_i$ the number of meals eaten by philosopher $i$. Throughput is
\[
r_{\text{thru}}=\frac{\sum_{i=1}^{N}m_i}{TN}.
\]
The denominator is the stated upper bound of one meal per philosopher per round; it does not account for turns spent acquiring forks and need not be attainable. The fairness penalty $r_{\text{fair}}$ is the variance of the meal counts across philosophers. The idle term $r_{\text{idle}}$ is the fraction of action attempts that invoke the no-op think tool; it does not measure all forms of unproductive behavior. We retain the weights used in the reported experiments: $\alpha=1.0$, $\beta=0.5$, $\gamma=0.3$, and $\delta=0.1$. Figure~\ref{fig:reward} summarizes the four terms.

\textbf{Reward without meals.}
If no philosopher eats, throughput and meal-count variance are zero. For such an episode, Equation~\ref{eq:reward} reduces to
\[
r=r_{\text{nd}}-0.1\,r_{\text{idle}}.
\]
If the episode also ends without deadlock and has zero idle penalty, its reward is $1.0$. Zero meals alone is insufficient for this conclusion: think calls can contribute an idle penalty, and deadlock removes the bonus. This example demonstrates that a high reward need not require productive behavior. We examine the corresponding reported model outcomes in Section~\ref{sec:discussion}.

\begin{figure}[!ht]
\centering
\begin{minipage}{0.96\linewidth}
\centering
\begin{tabular}{@{}cccc@{}}
\toprule
No-deadlock bonus & Throughput bonus & Fairness penalty & Idle penalty \\
$+1.0\,r_{\text{nd}}$ & $+0.5\,r_{\text{thru}}$ &
$-0.3\,r_{\text{fair}}$ & $-0.1\,r_{\text{idle}}$ \\
\bottomrule
\end{tabular}
\par\medskip
\fbox{\begin{minipage}{0.93\linewidth}
\centering
No meals: $r_{\text{thru}}=r_{\text{fair}}=0$.\\
With no deadlock: $r=1.0-0.1\,r_{\text{idle}}$.\\
With zero idle penalty as well: $r=1.0$.
\end{minipage}}
\end{minipage}
\caption{The reported reward weights and a zero-meal case. Zero meals eliminate the throughput and fairness terms, but do not determine the deadlock or idle terms. A reward of $1.0$ follows when deadlock and the idle penalty are also absent.}
\label{fig:reward}
\end{figure}

\section{Models and evaluation protocol}
\label{sec:method}

\textbf{Models.}
The main experiment compares seven model conditions. Claude Opus 4.7, GPT-5.4, Gemini 3.1 Pro, and Grok 4.2 are accessed through OpenRouter. Mistral-Small 24B is an open-weight comparator. Qwen3-14B~\cite{yang2025qwen3} is evaluated in its base form and with an adapter trained using GRPO~\cite{shao2024deepseekmath}. The separate experiment in Appendix~\ref{app:qwen8b} evaluates Qwen3-8B, Qwen3-8B+GRPO, and DeepSeek-R1-Distill-Qwen-7B~\cite{deepseekai2025r1}.

\begin{figure}[t]
\centering
\includegraphics[width=\linewidth]{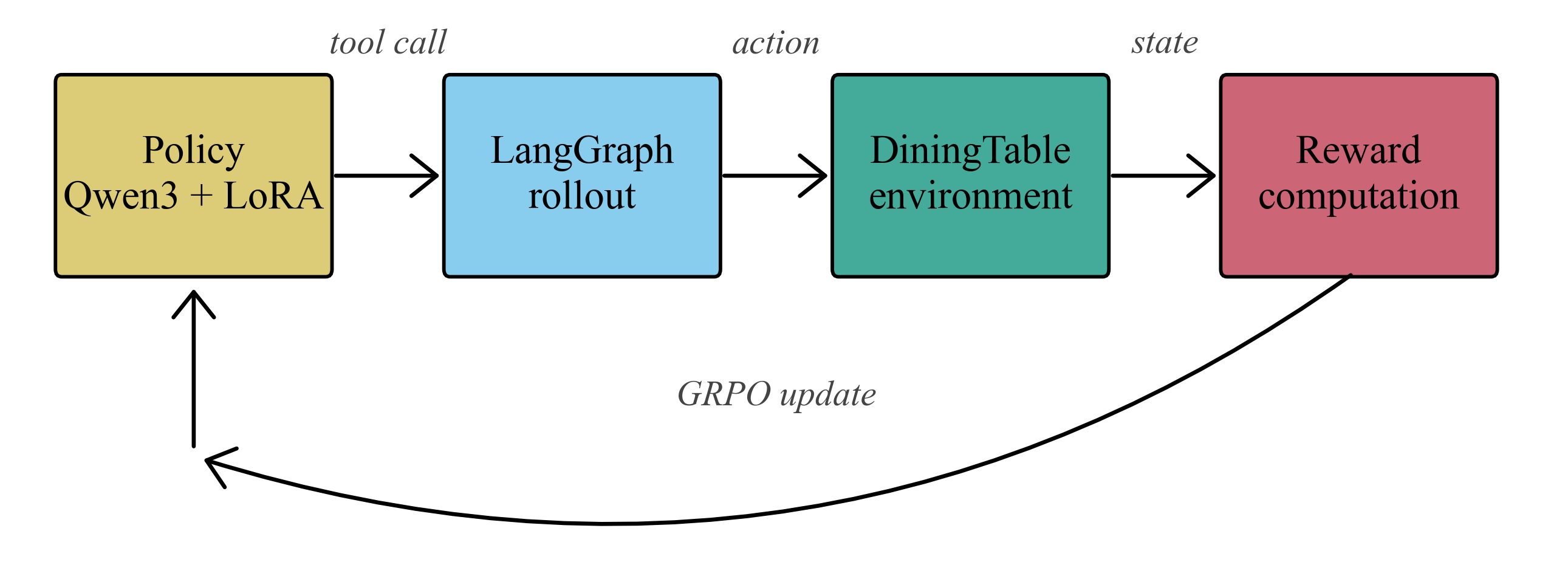}
\caption{Training pipeline for the Qwen3 policies. LangGraph mediates tool calls to the dining-table environment; completed rollouts receive the four-component reward used for GRPO updates to the LoRA adapter.}
\label{fig:pipeline}
\end{figure}

\textbf{Rollout pipeline.}
LangGraph wraps each philosopher's interaction with the policy model. On each turn, the model receives the system prompt and a message describing its adjacent forks and meal count, as specified in Appendix~\ref{app:reproducibility}. The environment applies the returned tool call and updates the state (Figure~\ref{fig:pipeline}). Open-weight models are served locally through vLLM~\cite{kwon2023vllm}, with the Hermes parser for Qwen3 and the Mistral parser for Mistral-Small 24B. Closed-source models use OpenRouter's tool-calling API. Transient failures are handled with exponential-backoff retries. The appendix gives the prompts, tool descriptions, and serving details.

\textbf{Training.}
We fine-tune Qwen3-14B and Qwen3-8B using GRPO~\cite{shao2024deepseekmath} in OpenPipe ART, with a rank-8 LoRA adapter~\cite{hu2022lora} and the reward in Equation~\ref{eq:reward}. Training rollouts use five philosophers and five rounds per episode. Each step samples two groups of three episodes, uses two epochs over the collected trajectories, and applies a single Adam~\cite{kingma2015adam} update. Each run comprises 16 steps. The reported configuration checkpoints the adapter after each step; the evaluation uses the saved step-15 adapter. Appendix~\ref{app:reproducibility} records the learning rate and other reported hyperparameters.

\textbf{Evaluation.}
Each of the seven main conditions is evaluated on 30 episodes at each of $N\in\{5,10,15\}$ with five rounds per episode: 90 episodes per condition and 630 in total. The smaller-scale experiment uses ten episodes per cell for three policies, $N\in\{5,6,7\}$, and round counts $\{5,10,15\}$, giving 270 additional episodes. Episode seeds depend only on $N$ and the episode index, so corresponding episodes across models use the same shuffled philosopher orderings. Training occurs at $N=5$; evaluation at other counts measures the saved policy's behavior at those scales.

\begin{samepage}
\section{Implementation and statistical analysis}
\label{sec:experiments}

\textbf{Hardware and serving.}
Training uses a single A100-SXM4 80\,GB GPU for the Qwen3-8B and Qwen3-14B runs, with OpenPipe ART and in-process vLLM serving at half GPU memory utilization. The reported evaluation setup uses four A100 80\,GB GPUs to serve four open-weight models in parallel, one per GPU. Closed-source models are accessed through OpenRouter's chat completion API.
\par
\end{samepage}

\textbf{Statistical analysis.}
All reported metrics use the same environment and the reward weights in Equation~\ref{eq:reward}. Confidence intervals use the bootstrap percentile method with 1000 resamples and seed 0, at the 95\,\% level. The reward comparisons between base and trained Qwen3-14B use Welch's $t$-tests on episode rewards and Hedges' $g$ effect sizes (Appendix~\ref{app:stats}). These comparisons concern the evaluated policies; the episodes are not independent training replications.

\textbf{Parser-dependent outcome.}
DeepSeek-R1-Distill-Qwen-7B is included only in the smaller-scale experiment. Its outputs yield no parseable tool calls under the Hermes serving configuration. We retain that observation to examine how the reward scores an episode without productive actions, with the serving limitation explicit in the interpretation.

\section{Results}
\label{sec:results}

\begin{table}[t]
\caption{Mean reward and diagnostic metrics at five philosophers across 30 episodes per model condition. Bold marks the highest reward and throughput and the lowest deadlock rate. These column extrema do not define an overall coordination ranking. Low fairness and idle penalties alone do not establish meal production.}
\label{tab:main-results}
\centering
\begin{tabular}{lccccc}
\toprule
Model & Reward & Deadlock & Throughput & Fairness penalty & Idle ratio \\
\midrule
Claude Opus 4.7 & \textbf{0.866} & \textbf{0.100} & \textbf{0.105} & 0.211 & 0.233 \\
GPT-5.4 & 0.454 & 0.533 & 0.057 & 0.101 & 0.113 \\
Gemini 3.1 Pro & 0.676 & 0.300 & 0.088 & 0.173 & 0.159 \\
Grok 4.2 & 0.848 & \textbf{0.100} & 0.076 & 0.203 & 0.292 \\
Mistral-Small 24B & 0.827 & 0.167 & 0.000 & 0.000 & 0.062 \\
Qwen3-14B & 0.128 & 0.867 & 0.020 & 0.032 & 0.053 \\
Qwen3-14B+GRPO & 0.092 & 0.900 & 0.008 & 0.024 & 0.044 \\
\bottomrule
\end{tabular}
\end{table}

\begin{figure}[!htbp]
\centering
\includegraphics[width=\linewidth]{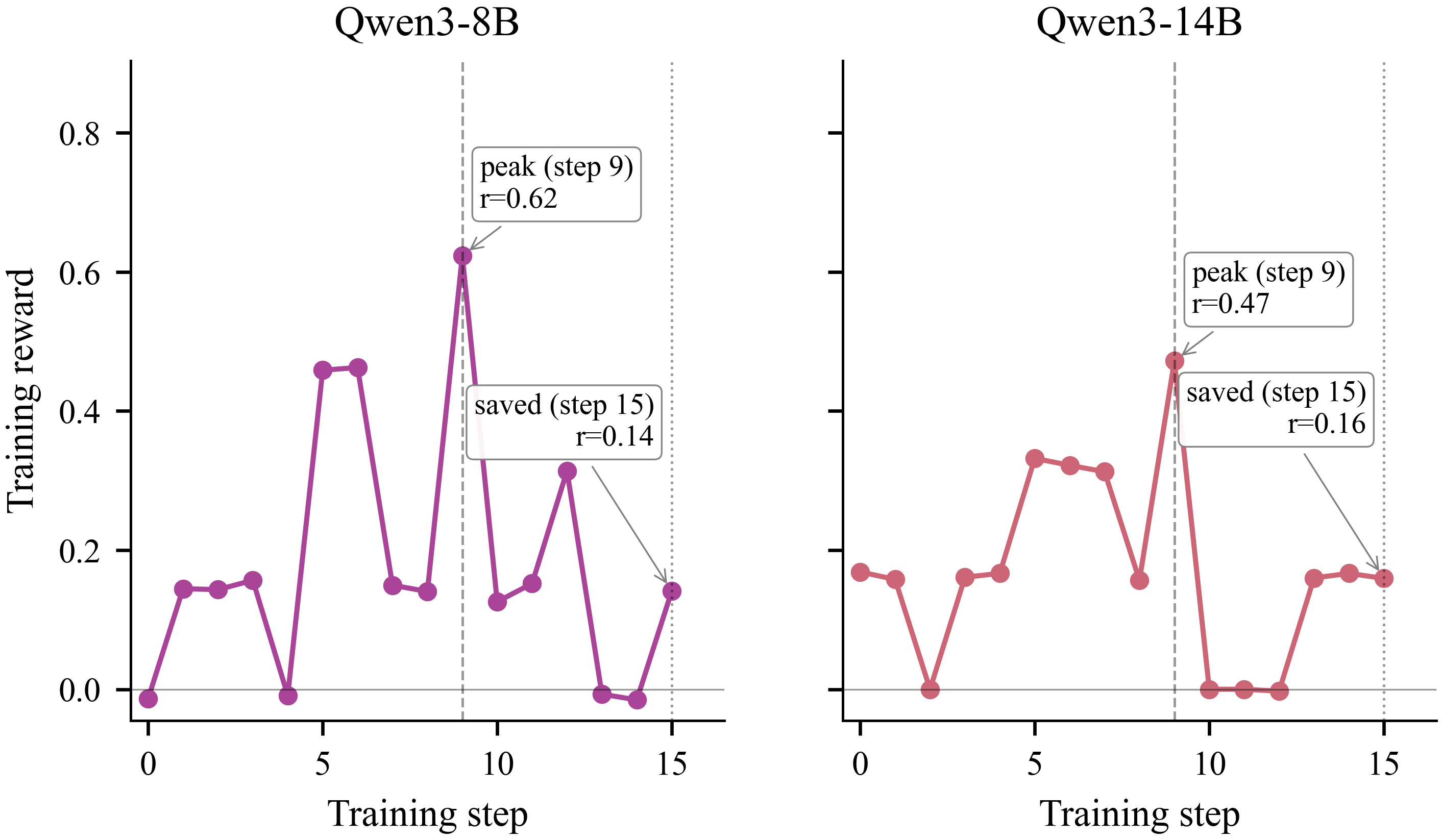}
\caption{Observed GRPO training rewards for the 8B and 14B runs. Both peak at step nine and are lower at step 15. The curves do not provide a benchmark evaluation of earlier checkpoints.}
\label{fig:training-curves}
\end{figure}

\begin{figure}[!htbp]
\centering
\includegraphics[width=\linewidth]{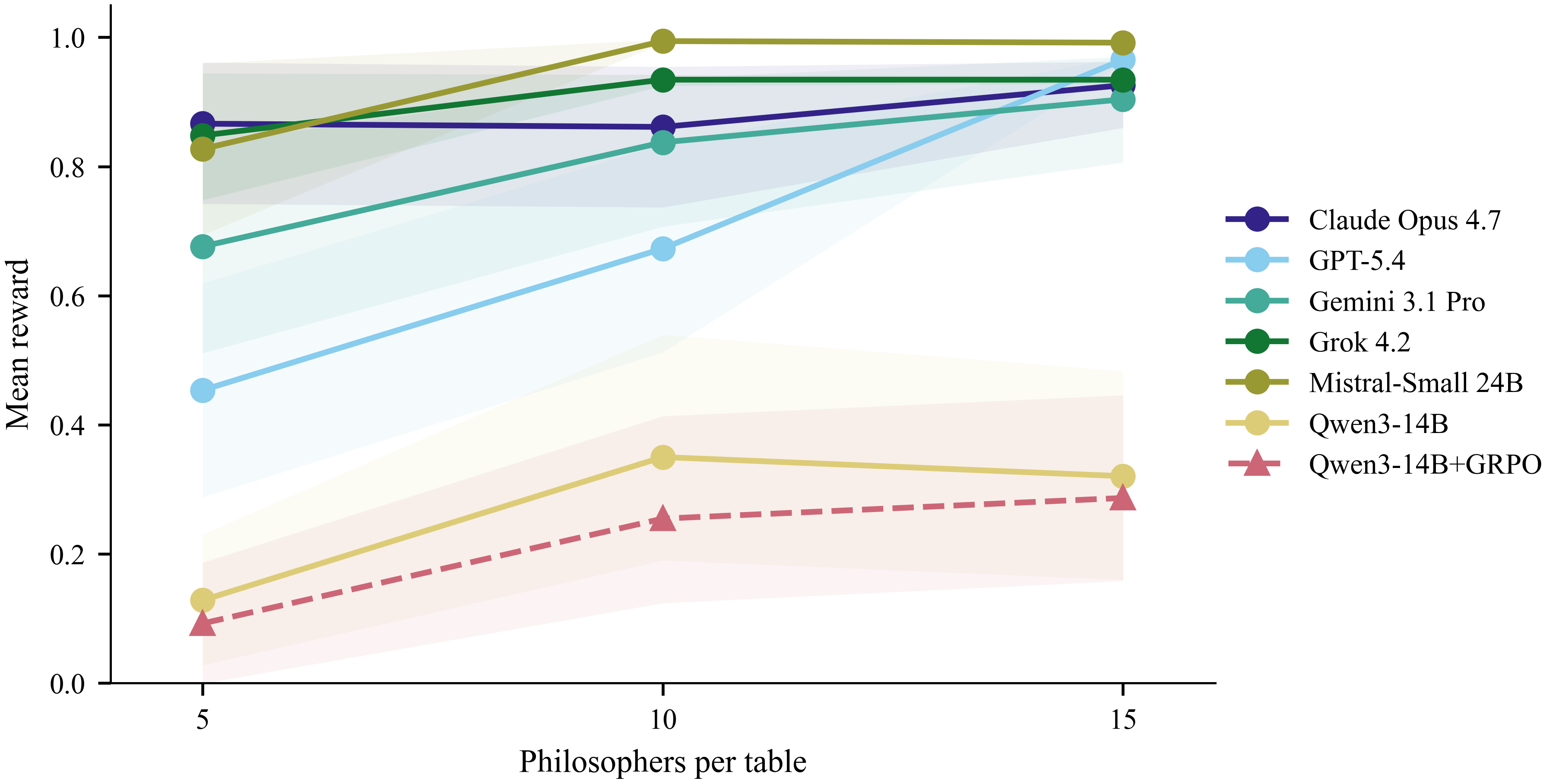}
\caption{Mean reward across five, ten, and fifteen philosophers. The trained Qwen3-14B policy has lower point estimates than its base at each count. Mistral-Small 24B has the highest mean reward at ten and fifteen philosophers; reward alone does not establish a ranking of productive coordination.}
\label{fig:scaling}
\end{figure}

\begingroup
\interlinepenalty=10000
\textbf{Reward and behavior at five philosophers.}
Table~\ref{tab:main-results} reports reward and its diagnostic components. The four closed-source models have mean rewards of 0.454--0.866, throughput of 0.057--0.105, and deadlock rates of 0.100--0.533. Mistral-Small 24B receives reward 0.827, within that reward range, but has zero throughput. Its deadlock rate is 0.167: avoiding deadlock in this setting does not imply that meals occur. Base Qwen3-14B has throughput 0.020 and deadlock rate 0.867; the trained policy has throughput 0.008 and deadlock rate 0.900. The closed-source models therefore have higher throughput and lower deadlock point estimates than either Qwen3-14B condition at this scale. Mistral's reward rank does not imply a corresponding rank in meal production.

\textbf{Reward across philosopher counts.}
Figure~\ref{fig:scaling} and Table~\ref{tab:scaling} report mean reward at $N\in\{5,10,15\}$. GPT-5.4's mean reward rises from 0.454 to 0.673 to 0.965; Claude Opus 4.7's values are 0.866, 0.861, and 0.926. Mistral-Small 24B has the highest mean reward in the table at ten and fifteen philosophers, at 0.994 and 0.992. These are scalar-reward comparisons, not a ranking of meal production across counts. The saved Qwen3-14B+GRPO policy has lower reward point estimates than its base at all three counts; we report the corresponding tests below.

\textbf{Reward comparison for the saved policy.}
Mean reward is 0.092 for the saved Qwen3-14B+GRPO policy and 0.128 for its base at five philosophers. The reported Welch test gives $t=-0.44$, $p=0.66$, and Hedges' $g=-0.11$. At ten and fifteen philosophers, $p=0.42$ and $p=0.77$; all three effect sizes satisfy $|g|<0.22$ (Appendix~\ref{app:stats}). None of these tests rejects equal mean reward at $\alpha=0.05$. This result does not establish equivalence or describe variability across training runs. The lower throughput and higher deadlock at five philosophers are descriptive point estimates; the reported tests concern reward.

\textbf{Training reward and checkpoint evaluation.}
Both training curves peak at step nine: the reported training rewards are 0.62 for Qwen3-8B and 0.47 for Qwen3-14B. At step 15, they are 0.14 and 0.16, respectively (Figure~\ref{fig:training-curves}). These values are observations from training, with six on-policy rollouts collected per step. Their decline does not by itself determine how the earlier adapters would perform on the evaluation benchmark. The reported evaluation compares the saved step-15 policy with its base, rather than comparing checkpoints under a common evaluation protocol.

\endgroup

\begin{table}[!htbp]
\caption{Mean reward for the seven main model conditions at five, ten, and fifteen philosophers. Bold marks the largest mean reward in each column.}
\label{tab:scaling}
\centering
\begin{tabular}{lccc}
\toprule
& \multicolumn{3}{c}{Philosophers} \\
\cmidrule(lr){2-4}
Model & 5 & 10 & 15 \\
\midrule
Claude Opus 4.7 & \textbf{0.866} & 0.861 & 0.926 \\
GPT-5.4 & 0.454 & 0.673 & 0.965 \\
Gemini 3.1 Pro & 0.676 & 0.837 & 0.904 \\
Grok 4.2 & 0.848 & 0.934 & 0.934 \\
Mistral-Small 24B & 0.827 & \textbf{0.994} & \textbf{0.992} \\
Qwen3-14B & 0.128 & 0.350 & 0.320 \\
Qwen3-14B+GRPO & 0.092 & 0.255 & 0.287 \\
\bottomrule
\end{tabular}
\end{table}

\FloatBarrier
\section{Discussion}
\label{sec:discussion}

\begin{figure}[!htbp]
\centering
\includegraphics[width=\linewidth]{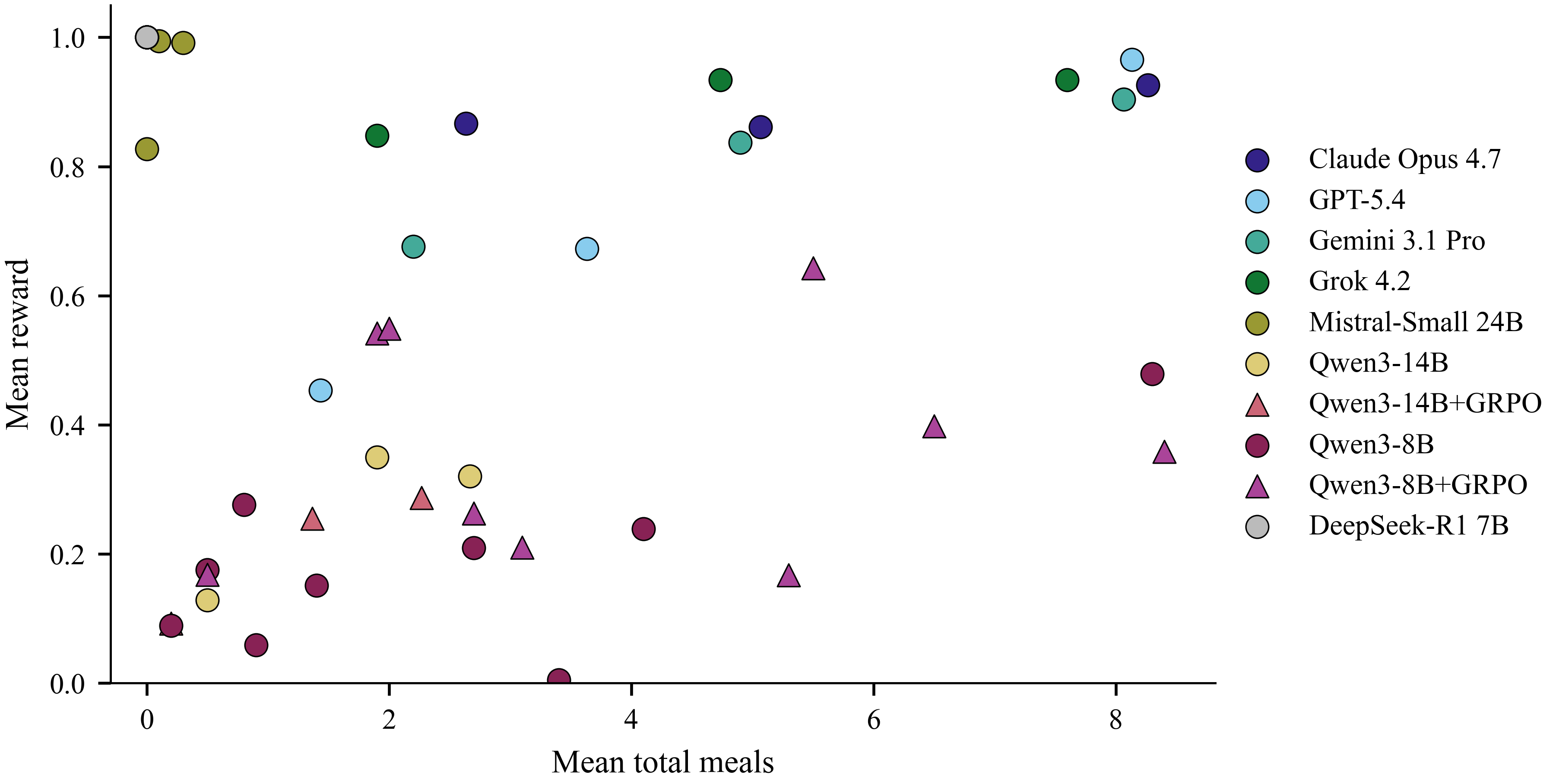}
\caption{Mean reward and mean total meals for the evaluated model, philosopher-count, and round-budget conditions. Total meals are not normalized across the differing counts and budgets; this plot illustrates reward without meals rather than a cross-scale performance ranking. DeepSeek-R1-Distill-Qwen-7B receives mean reward 1.0 with zero meals in all nine smaller-scale conditions under its reported parser limitation. Mistral-Small 24B receives mean reward 0.827 with zero meals at five philosophers. High reward need not indicate meal production.}
\label{fig:reward-vs-meals}
\end{figure}

\textbf{Reward without productive behavior.}
Figure~\ref{fig:reward-vs-meals} compares mean reward with mean total meals across the reported evaluation conditions. Mistral-Small 24B receives mean reward 0.827 with zero meals at five philosophers. DeepSeek-R1-Distill-Qwen-7B receives mean reward 1.0 with zero meals in all nine conditions of the smaller-scale experiment, where its outputs yield no parseable tool calls under the Hermes serving configuration. These observations have different operational contexts, but both show why reward alone is insufficient to establish productive coordination. The DeepSeek result also depends on the serving configuration; it is not evidence of a general inability of that model to coordinate.

The reward definition explains how such outcomes are possible. The no-deadlock bonus does not require a meal, and zero meals eliminate both the throughput and meal-count variance terms. Reward is then determined by deadlock and the think-action penalty, as derived in Section~\ref{sec:task}. Consequently, a low fairness penalty can reflect uniformly zero meals rather than equitable productive behavior, and a low idle ratio does not establish progress. Conditioning the no-deadlock bonus on meal production or using throughput as the primary training signal are possible changes, not interventions evaluated here. Their effects on learned behavior remain unknown.

\textbf{Interpreting episode rewards.}
The per-episode distributions in Appendix~\ref{app:supplementary} complement the mean rewards by showing their spread. Their concentrations near zero and one cannot be labeled simply as failure and success: an episode with reward near one may contain no meals. Mean reward therefore does not measure the fraction of episodes with productive coordination.

\textbf{Interpreting the saved-policy comparison.}
The saved-policy comparison illustrates why reward tests and behavioral measures answer different questions. The tests concern mean scalar reward; throughput and deadlock describe task outcomes. A nonsignificant reward difference therefore does not establish equivalent behavior. The lower throughput and higher deadlock point estimates at five philosophers characterize the evaluated checkpoint, while their cause remains unresolved (Section~\ref{sec:limitations}).

Checkpoint selection raises a related measurement issue: selecting the largest observed training reward would still select on an on-policy training signal, not on held-out performance. The earlier peaks motivate a common evaluation of checkpoints; they do not identify a better-performing policy.

\FloatBarrier
\section{Limitations}
\label{sec:limitations}

\textbf{Scope of the training comparison.}
The main comparison uses one Qwen3-14B training run and one evaluated saved checkpoint; the separate Qwen3-8B experiment is not a replication of that run. Evaluation episodes characterize these policies, not variability across independently trained policies, so they cannot establish a general GRPO effect. No common evaluation of earlier checkpoints is reported, so the results do not identify the best checkpoint. The reward tests do not establish policy equivalence, and the throughput and deadlock comparisons remain descriptive point estimates.

\textbf{Reward design and causal interpretation.}
The zero-meal examples establish a limitation of the evaluated reward, not of scalar rewards in general. Training uses this same reward, but the study does not vary the reward design or training procedure to isolate the cause of the saved policy's behavior. Consequently, the results do not establish that the reward's allowance for zero-meal outcomes caused that behavior or that a revised reward would improve coordination.

\textbf{Benchmark scope and calibration.}
The evaluation uses instances of the same model, shared prompts and tools, and specified turn schedules and episode budgets. Its results need not transfer to heterogeneous agents or other coordination tasks. No non-LLM policy baseline is included, limiting calibration of model performance against simple resource-allocation policies. This does not change the reported coexistence of high reward and zero meals, but it restricts broader claims about coordination capability.

\textbf{Serving dependence.}
Outcomes reflect both the policies and their serving interfaces. In particular, DeepSeek-R1-Distill-Qwen-7B yields no parseable tool calls under the reported configuration. Its result illustrates how the reward scores that operational failure, not an intrinsic inability of the model to coordinate. Cross-model comparisons should therefore be interpreted within the evaluated configurations.

\section{Conclusion}
\label{sec:conclusion}

In the evaluated dining-philosophers environment, high scalar reward is not sufficient evidence of productive coordination. The observed zero-meal outcomes and the reward decomposition identify the same issue: the no-deadlock bonus can reward an episode without progress. Reporting meal throughput and deadlock separately makes this distinction visible instead of treating reward rankings as rankings of coordination capability.

The Qwen3-14B comparison extends this measurement perspective to policy adaptation. The final checkpoint from one GRPO training run has lower throughput and higher deadlock than its base at five philosophers in point estimates, while reward tests detect no significant difference at any evaluated count. This result characterizes the evaluated checkpoint, not GRPO in general. Testing progress-conditioned rewards, evaluating checkpoints on held-out episodes, and repeating training runs would help determine whether changes to the training procedure improve task progress. The present evidence supports assessing that progress directly alongside the reward used for training and evaluation.

\bibliographystyle{plainnat}
\bibliography{references}

@article{dijkstra1971hierarchical,
  author    = {Dijkstra, Edsger W.},
  title     = {Hierarchical ordering of sequential processes},
  journal   = {Acta Informatica},
  volume    = {1},
  number    = {2},
  pages     = {115--138},
  year      = {1971},
  publisher = {Springer},
  doi       = {10.1007/BF00289519}
}

@inproceedings{kingma2015adam,
  author    = {Kingma, Diederik P. and Ba, Jimmy},
  title     = {{Adam}: A Method for Stochastic Optimization},
  booktitle = {International Conference on Learning Representations (ICLR)},
  year      = {2015}
}

@article{schulman2017ppo,
  author  = {Schulman, John and Wolski, Filip and Dhariwal, Prafulla and Radford, Alec and Klimov, Oleg},
  title   = {Proximal Policy Optimization Algorithms},
  journal = {arXiv preprint arXiv:1707.06347},
  year    = {2017}
}

@inproceedings{christiano2017rlhf,
  author    = {Christiano, Paul and Leike, Jan and Brown, Tom B. and Martic, Miljan and Legg, Shane and Amodei, Dario},
  title     = {Deep Reinforcement Learning from Human Preferences},
  booktitle = {Advances in Neural Information Processing Systems (NeurIPS)},
  year      = {2017}
}

@article{bard2020hanabi,
  author  = {Bard, Nolan and Foerster, Jakob N. and Chandar, Sarath and Burch, Neil and Lanctot, Marc and Song, H. Francis and Parisotto, Emilio and Dumoulin, Vincent and Moitra, Subhodeep and Hughes, Edward and Dunning, Iain and Mourad, Shibl and Larochelle, Hugo and Bellemare, Marc G. and Bowling, Michael},
  title   = {The {Hanabi} Challenge: A New Frontier for {AI} Research},
  journal = {Artificial Intelligence},
  volume  = {280},
  pages   = {103216},
  year    = {2020},
  publisher = {Elsevier}
}

@inproceedings{carroll2019overcooked,
  author    = {Carroll, Micah and Shah, Rohin and Ho, Mark K. and Griffiths, Thomas L. and Seshia, Sanjit A. and Abbeel, Pieter and Dragan, Anca},
  title     = {On the Utility of Learning about Humans for Human-{AI} Coordination},
  booktitle = {Advances in Neural Information Processing Systems (NeurIPS)},
  year      = {2019}
}

@article{meta2022cicero,
  author  = {{Meta Fundamental AI Research Diplomacy Team (FAIR)} and Bakhtin, Anton and Brown, Noam and Dinan, Emily and Farina, Gabriele and Flaherty, Colin and Fried, Daniel and Goff, Andrew and Gray, Jonathan and Hu, Hengyuan and others},
  title   = {Human-level play in the game of {Diplomacy} by combining language models with strategic reasoning},
  journal = {Science},
  volume  = {378},
  number  = {6624},
  pages   = {1067--1074},
  year    = {2022},
  publisher = {American Association for the Advancement of Science},
  doi     = {10.1126/science.ade9097}
}

@inproceedings{hu2022lora,
  author    = {Hu, Edward J. and Shen, Yelong and Wallis, Phillip and Allen{-}Zhu, Zeyuan and Li, Yuanzhi and Wang, Shean and Wang, Lu and Chen, Weizhu},
  title     = {{LoRA}: Low-Rank Adaptation of Large Language Models},
  booktitle = {International Conference on Learning Representations (ICLR)},
  year      = {2022},
  url       = {https://openreview.net/forum?id=nZeVKeeFYf9}
}

@inproceedings{kwon2023vllm,
  author    = {Kwon, Woosuk and Li, Zhuohan and Zhuang, Siyuan and Sheng, Ying and Zheng, Lianmin and Yu, Cody Hao and Gonzalez, Joseph E. and Zhang, Hao and Stoica, Ion},
  title     = {Efficient Memory Management for Large Language Model Serving with {PagedAttention}},
  booktitle = {Proceedings of the 29th Symposium on Operating Systems Principles (SOSP)},
  year      = {2023},
  publisher = {Association for Computing Machinery},
  doi       = {10.1145/3600006.3613165}
}

@inproceedings{ouyang2022instructgpt,
  author    = {Ouyang, Long and Wu, Jeffrey and Jiang, Xu and Almeida, Diogo and Wainwright, Carroll L. and Mishkin, Pamela and Zhang, Chong and Agarwal, Sandhini and Slama, Katarina and Ray, Alex and Schulman, John and Hilton, Jacob and Kelton, Fraser and Miller, Luke and Simens, Maddie and Askell, Amanda and Welinder, Peter and Christiano, Paul and Leike, Jan and Lowe, Ryan},
  title     = {Training Language Models to Follow Instructions with Human Feedback},
  booktitle = {Advances in Neural Information Processing Systems (NeurIPS)},
  year      = {2022}
}

@inproceedings{yao2023react,
  author    = {Yao, Shunyu and Zhao, Jeffrey and Yu, Dian and Du, Nan and Shafran, Izhak and Narasimhan, Karthik and Cao, Yuan},
  title     = {{ReAct}: Synergizing Reasoning and Acting in Language Models},
  booktitle = {International Conference on Learning Representations (ICLR)},
  year      = {2023}
}

@inproceedings{schick2023toolformer,
  author    = {Schick, Timo and Dwivedi-Yu, Jane and Dess{\`\i}, Roberto and Raileanu, Roberta and Lomeli, Maria and Hambro, Eric and Zettlemoyer, Luke and Cancedda, Nicola and Scialom, Thomas},
  title     = {{Toolformer}: Language Models Can Teach Themselves to Use Tools},
  booktitle = {Advances in Neural Information Processing Systems (NeurIPS)},
  year      = {2023}
}

@article{liang2023helm,
  author  = {Liang, Percy and Bommasani, Rishi and Lee, Tony and Tsipras, Dimitris and Soylu, Dilara and Yasunaga, Michihiro and Zhang, Yian and Narayanan, Deepak and Wu, Yuhuai and Kumar, Ananya and others},
  title   = {Holistic Evaluation of Language Models},
  journal = {Transactions on Machine Learning Research},
  year    = {2023},
  url     = {https://openreview.net/forum?id=iO4LZibEqW}
}

@inproceedings{mialon2024gaia,
  author    = {Mialon, Gr{\'e}goire and Fourrier, Cl{\'e}mentine and Swift, Craig and Wolf, Thomas and LeCun, Yann and Scialom, Thomas},
  title     = {{GAIA}: a benchmark for General {AI} Assistants},
  booktitle = {International Conference on Learning Representations (ICLR)},
  year      = {2024}
}

@inproceedings{yao2025taubench,
  author    = {Yao, Shunyu and Shinn, Noah and Razavi, Pedram and Narasimhan, Karthik},
  title     = {{$\tau$}-bench: A Benchmark for Tool-Agent-User Interaction in Real-World Domains},
  booktitle = {International Conference on Learning Representations (ICLR)},
  year      = {2025}
}

@inproceedings{liu2024agentbench,
  author    = {Liu, Xiao and Yu, Hao and Zhang, Hanchen and Xu, Yifan and Lei, Xuanyu and Lai, Hanyu and Gu, Yu and Ding, Hangliang and Men, Kaiwen and Yang, Kejuan and Zhang, Shudan and Deng, Xiang and Zeng, Aohan and Du, Zhengxiao and Zhang, Chenhui and Shen, Sheng and Zhang, Tianjun and Su, Yu and Sun, Huan and Huang, Minlie and Dong, Yuxiao and Tang, Jie},
  title     = {{AgentBench}: Evaluating {LLMs} as Agents},
  booktitle = {International Conference on Learning Representations (ICLR)},
  year      = {2024}
}

@inproceedings{zhou2024webarena,
  author    = {Zhou, Shuyan and Xu, Frank F. and Zhu, Hao and Zhou, Xuhui and Lo, Robert and Sridhar, Abishek and Cheng, Xianyi and Ou, Tianyue and Bisk, Yonatan and Fried, Daniel and Alon, Uri and Neubig, Graham},
  title     = {{WebArena}: A Realistic Web Environment for Building Autonomous Agents},
  booktitle = {International Conference on Learning Representations (ICLR)},
  year      = {2024}
}

@inproceedings{jimenez2024swebench,
  author    = {Jimenez, Carlos E. and Yang, John and Wettig, Alexander and Yao, Shunyu and Pei, Kexin and Press, Ofir and Narasimhan, Karthik R.},
  title     = {{SWE}-bench: Can Language Models Resolve Real-World {GitHub} Issues?},
  booktitle = {International Conference on Learning Representations (ICLR)},
  year      = {2024}
}

@inproceedings{li2023camel,
  author    = {Li, Guohao and Hammoud, Hasan Abed Al Kader and Itani, Hani and Khizbullin, Dmitrii and Ghanem, Bernard},
  title     = {{CAMEL}: Communicative Agents for ``Mind'' Exploration of Large Language Model Society},
  booktitle = {Advances in Neural Information Processing Systems (NeurIPS)},
  year      = {2023}
}

@inproceedings{hong2024metagpt,
  author    = {Hong, Sirui and Zhuge, Mingchen and Chen, Jonathan and Zheng, Xiawu and Cheng, Yuheng and Wang, Jinlin and Zhang, Ceyao and Wang, Zili and Yau, Steven Ka Shing and Lin, Zijuan and Zhou, Liyang and Ran, Chenyu and Xiao, Lingfeng and Wu, Chenglin and Schmidhuber, J{\"u}rgen},
  title     = {{MetaGPT}: Meta Programming for a Multi-Agent Collaborative Framework},
  booktitle = {International Conference on Learning Representations (ICLR)},
  year      = {2024}
}

@inproceedings{park2023generative,
  author    = {Park, Joon Sung and O'Brien, Joseph C. and Cai, Carrie J. and Morris, Meredith Ringel and Liang, Percy and Bernstein, Michael S.},
  title     = {Generative Agents: Interactive Simulacra of Human Behavior},
  booktitle = {Proceedings of the 36th Annual ACM Symposium on User Interface Software and Technology (UIST)},
  year      = {2023},
  doi       = {10.1145/3586183.3606763}
}

@inproceedings{wu2024autogen,
  author    = {Wu, Qingyun and Bansal, Gagan and Zhang, Jieyu and Wu, Yiran and Li, Beibin and Zhu, Erkang and Jiang, Li and Zhang, Xiaoyun and Zhang, Shaokun and Liu, Jiale and Awadallah, Ahmed Hassan and White, Ryen W. and Burger, Doug and Wang, Chi},
  title     = {{AutoGen}: Enabling Next-Gen {LLM} Applications via Multi-Agent Conversations},
  booktitle = {Conference on Language Modeling (COLM)},
  year      = {2024}
}

@inproceedings{du2024multiagent,
  author    = {Du, Yilun and Li, Shuang and Torralba, Antonio and Tenenbaum, Joshua B. and Mordatch, Igor},
  title     = {Improving Factuality and Reasoning in Language Models through Multiagent Debate},
  booktitle = {International Conference on Machine Learning (ICML)},
  year      = {2024}
}

@article{shao2024deepseekmath,
  author  = {Shao, Zhihong and Wang, Peiyi and Zhu, Qihao and Xu, Runxin and Song, Junxiao and Bi, Xiao and Zhang, Haowei and Zhang, Mingchuan and Li, Y.~K. and Wu, Y. and Guo, Daya},
  title   = {{DeepSeekMath}: Pushing the Limits of Mathematical Reasoning in Open Language Models},
  journal = {arXiv preprint arXiv:2402.03300},
  year    = {2024}
}

@article{yang2025qwen3,
  author  = {Yang, An and Li, Anfeng and Yang, Baosong and Zhang, Beichen and Hui, Binyuan and Zheng, Bo and Yu, Bowen and Gao, Chang and Huang, Chengen and Lv, Chenxu and Zheng, Chujie and Liu, Dayiheng and others},
  title   = {{Qwen3} Technical Report},
  journal = {arXiv preprint arXiv:2505.09388},
  year    = {2025}
}

@article{deepseekai2025r1,
  author  = {{DeepSeek-AI}},
  title   = {{DeepSeek-R1}: Incentivizing Reasoning Capability in {LLMs} via Reinforcement Learning},
  journal = {arXiv preprint arXiv:2501.12948},
  year    = {2025}
}

\appendix

\newpage

\section{Reproducibility details}
\label{app:reproducibility}

\textbf{GRPO hyperparameters.}
Both training runs use learning rate $1\times 10^{-5}$, three rollouts per group, two groups per training step, two epochs over each step's collected trajectories, and Adam~\cite{kingma2015adam} with default $\beta_1, \beta_2$. We run for sixteen steps total. The LoRA adapter has rank $r = 8$. Rollouts are collected at $N = 5$ philosophers and 5 rounds per episode.

\textbf{Reward weights.}
Equation~\ref{eq:reward} uses $\alpha = 1.0$, $\beta = 0.5$, $\gamma = 0.3$, $\delta = 0.1$. These are the values used during training and during all evaluation runs reported in this paper.

\textbf{Tool definitions.}
Each philosopher exposes six tools to the LLM: pick-up-left, pick-up-right, put-down-left, put-down-right, eat, and think. The first four manipulate fork ownership; eat is admissible only when both adjacent forks are held; think is a no-op. Tool docstrings (the strings the LLM sees) and the exact prompt text are reproduced verbatim below.

\textbf{Serving and operational details.}
Open-weight models are served locally with vLLM (Hermes tool-call parser for the Qwen3 family and DeepSeek-R1, Mistral parser for Mistral-Small 24B). Closed-source models are queried through OpenRouter's chat completion API with tool-calling enabled. Retry policy, episode-seed hashing, and the exact OpenRouter model identifiers are recorded verbatim in the released code repository.

\textbf{Prompts and tool schema.}
The reported setup uses the same system-prompt template, per-turn message template, and six tool descriptions across models and runs. The templates are instantiated with the philosopher identifiers and current state, as shown below. The reward formula in Equation~\ref{eq:reward} is never shown to the model; it is computed by the environment after each episode terminates and enters training only as a GRPO advantage signal.

\begin{figure}[!ht]
\begin{promptbox}
\small
You are Philosopher \{i\} at a circular table with \{n\} philosophers and \{n\} forks. Fork F\{left\} is on your left, fork F\{right\} is on your right. You need BOTH forks to eat. After eating, both forks are released. Goal: eat as many meals as possible. If all philosophers hold one fork each, deadlock occurs and nobody eats. Choose ONE action per turn.
\end{promptbox}
\caption{System prompt, formatted per philosopher with the philosopher index, total philosopher count, and adjacent fork indices substituted into the placeholders.}
\label{fig:prompt-system}
\end{figure}

\begin{figure}[!ht]
\begin{promptbox}
\small
You are Philosopher \{pid\}. Left fork (F\{lf\}): \{available $|$ held by P$x$\}. Right fork (F\{rf\}): \{available $|$ held by P$x$\}. Your meals eaten: \{count\}.
\end{promptbox}
\caption{Per-turn human message, formatted from the current table state immediately before the philosopher's turn.}
\label{fig:prompt-human}
\end{figure}

\begin{figure}[!ht]
\begin{promptbox}
\small
\textbf{pick\_up\_left:} Pick up the fork on your left.\\
\textbf{pick\_up\_right:} Pick up the fork on your right.\\
\textbf{put\_down\_left:} Put down the fork on your left.\\
\textbf{put\_down\_right:} Put down the fork on your right.\\
\textbf{eat:} Eat a meal. Requires both forks. Releases both after.\\
\textbf{think:} Wait this turn. Use to avoid deadlock.
\end{promptbox}
\caption{Tool descriptions exposed to the model in the tool schema; the LLM sees these strings as part of each tool's signature when deciding which action to take.}
\label{fig:prompt-tools}
\end{figure}

\newpage
\FloatBarrier
\section{Method detail figures}
\label{app:method}

\begin{figure}[!ht]
\centering
\includegraphics[width=\linewidth]{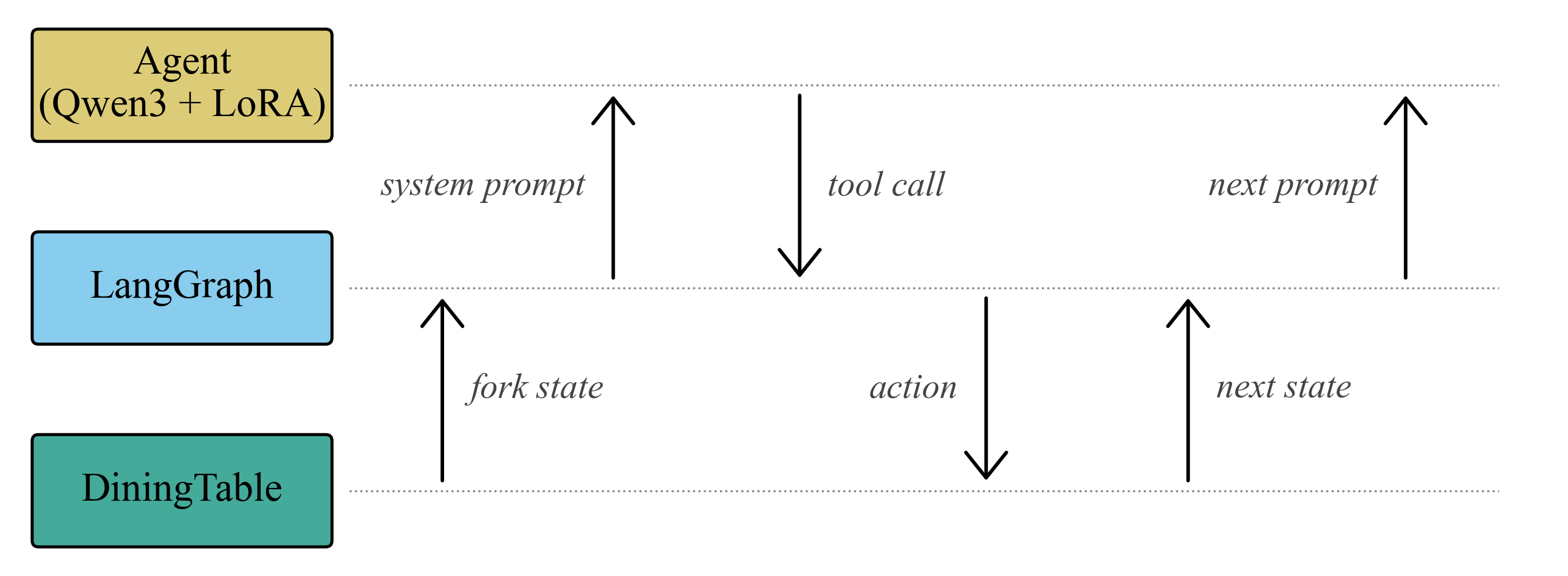}
\caption{Per-turn interaction, illustrated with the Qwen3+LoRA policy. LangGraph passes the system prompt and per-turn state message to the agent, and the environment executes its returned tool call. The prompt templates are shown in Figures~\ref{fig:prompt-system} and~\ref{fig:prompt-human}.}
\label{fig:turn}
\end{figure}

Figure~\ref{fig:turn} expands the rollout pipeline from Section~\ref{sec:method} into a per-turn swimlane, showing the data flow between the dining table, the LangGraph runtime, and the LLM agent on each turn.

\FloatBarrier
\section{Supplementary results}
\label{app:supplementary}

\begin{figure}[!ht]
\centering
\includegraphics[width=\linewidth]{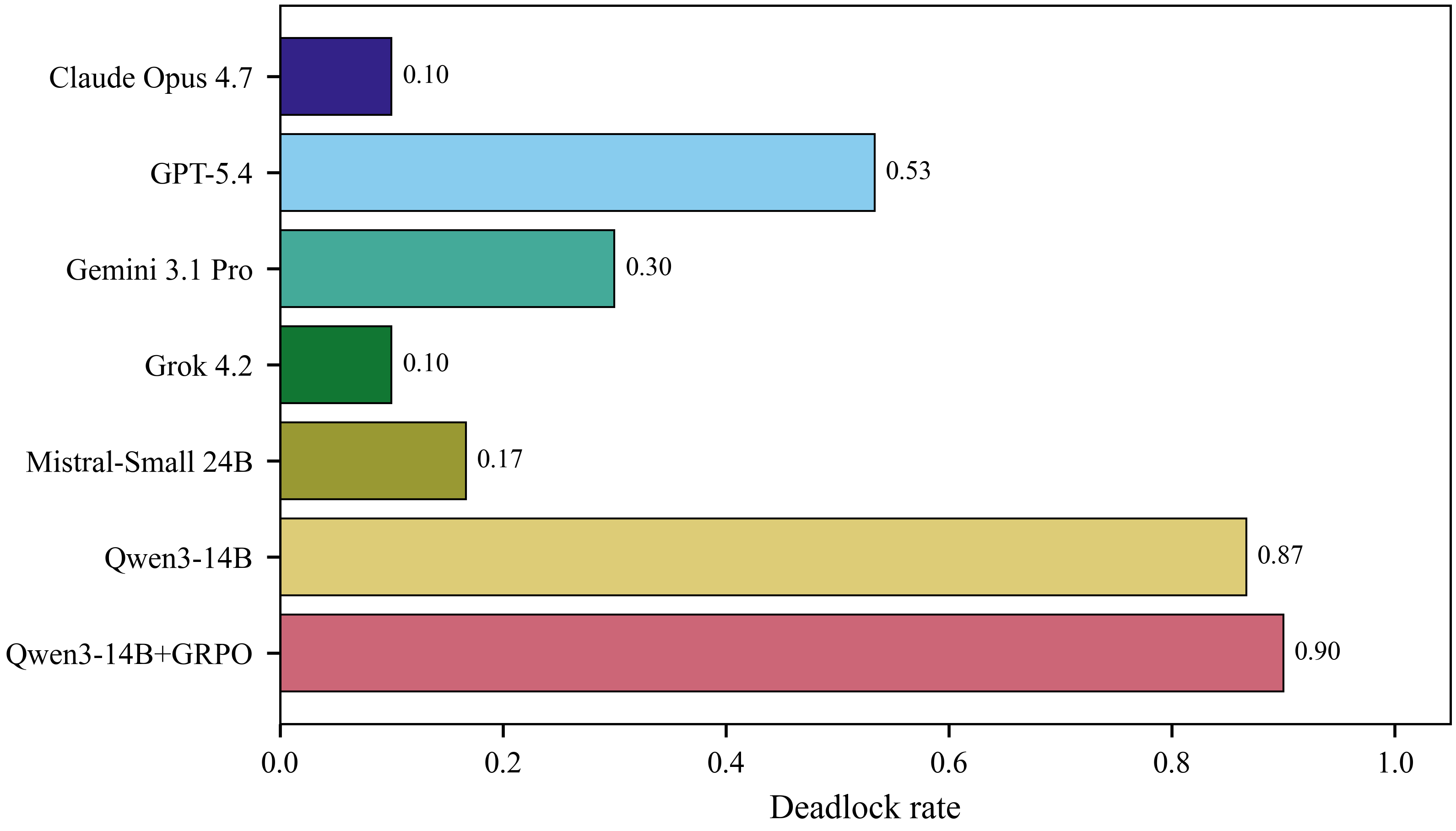}
\caption{Qwen3-14B and Qwen3-14B+GRPO deadlock in 87 to 90 percent of episodes at 5 philosophers, while the four closed-source models and Mistral-Small 24B stay below 54 percent.}
\label{fig:deadlock}
\end{figure}

\begin{figure}[!ht]
\centering
\includegraphics[width=\linewidth]{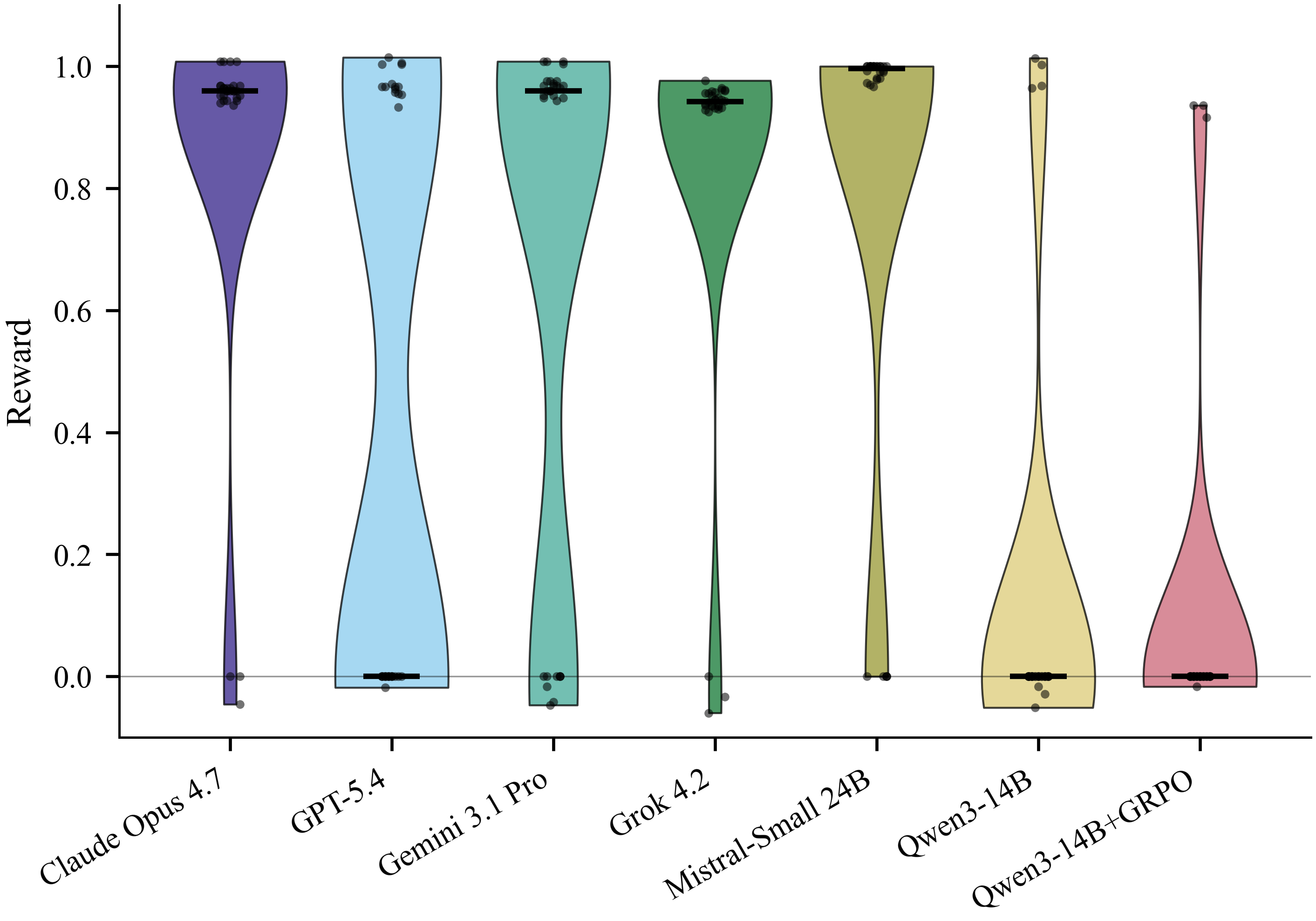}
\caption{Per-episode reward distributions at five philosophers for the seven main model conditions. Reward observations concentrate near zero and near one. High reward is not a success indicator, since episodes without meals can also receive high reward.}
\label{fig:distribution}
\end{figure}

Figure~\ref{fig:deadlock} reports deadlock rates for the seven main model conditions at five philosophers. These are behavioral measurements, distinct from scalar reward. Figure~\ref{fig:distribution} shows the corresponding per-episode reward distributions. Observations concentrate near zero and near one, but these reward ranges are not labels for failed and successful coordination: a high reward can occur without any meals. The distributions show variation hidden by the means; deadlock and throughput are needed to interpret that variation.

\FloatBarrier
\section{Smaller-scale Qwen3-8B experiment}
\label{app:qwen8b}

We ran a smaller-scale precursor experiment to the main Qwen3-14B study using Qwen3-8B, Qwen3-8B+GRPO, and DeepSeek-R1-Distill-Qwen-7B as a third open-weight baseline. The philosopher counts are $\{5, 6, 7\}$ instead of $\{5, 10, 15\}$, and the rounds-per-episode setting is varied across $\{5, 10, 15\}$ rather than fixed at five. Each (model, philosopher count, rounds) cell uses ten episodes, for $3 \times 3 \times 3 \times 10 = 270$ episodes total.

The direction of the saved Qwen3-8B policy's reward difference from its base varies across evaluation settings. At five philosophers and five rounds, mean reward is 0.54 for Qwen3-8B+GRPO and 0.09 for its base. At five philosophers and fifteen rounds, the ordering reverses: 0.17 for the trained policy and 0.24 for its base. At six philosophers and five rounds, the means are 0.17 and 0.18, respectively. These are descriptive reward comparisons, not evidence of a consistent coordination improvement or of variability across training seeds. DeepSeek-R1-Distill-Qwen-7B receives mean reward 1.0 with zero meals in all nine conditions; its outputs yield no parseable tool calls under the reported serving configuration.

\begin{figure}[!ht]
\centering
\includegraphics[width=0.95\linewidth]{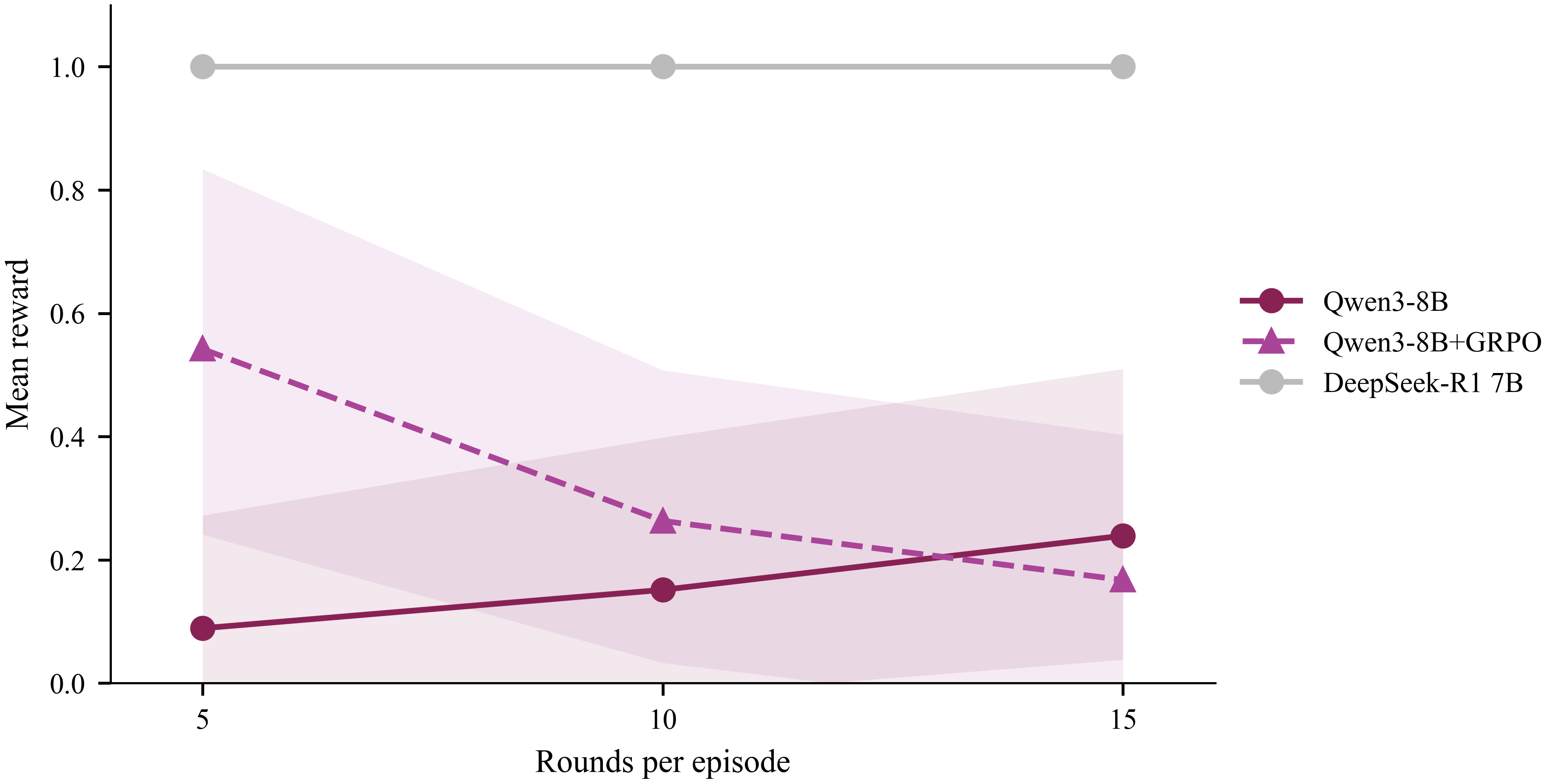}
\caption{Mean reward at five philosophers as the episode budget varies. Qwen3-8B+GRPO has higher mean reward than its base at five rounds (0.54 versus 0.09), but lower mean reward at fifteen rounds (0.17 versus 0.24). DeepSeek-R1-Distill-Qwen-7B receives mean reward 1.0 under the reported parser limitation.}
\label{fig:qwen8b-round}
\end{figure}

\begin{figure}[!ht]
\centering
\includegraphics[width=0.95\linewidth]{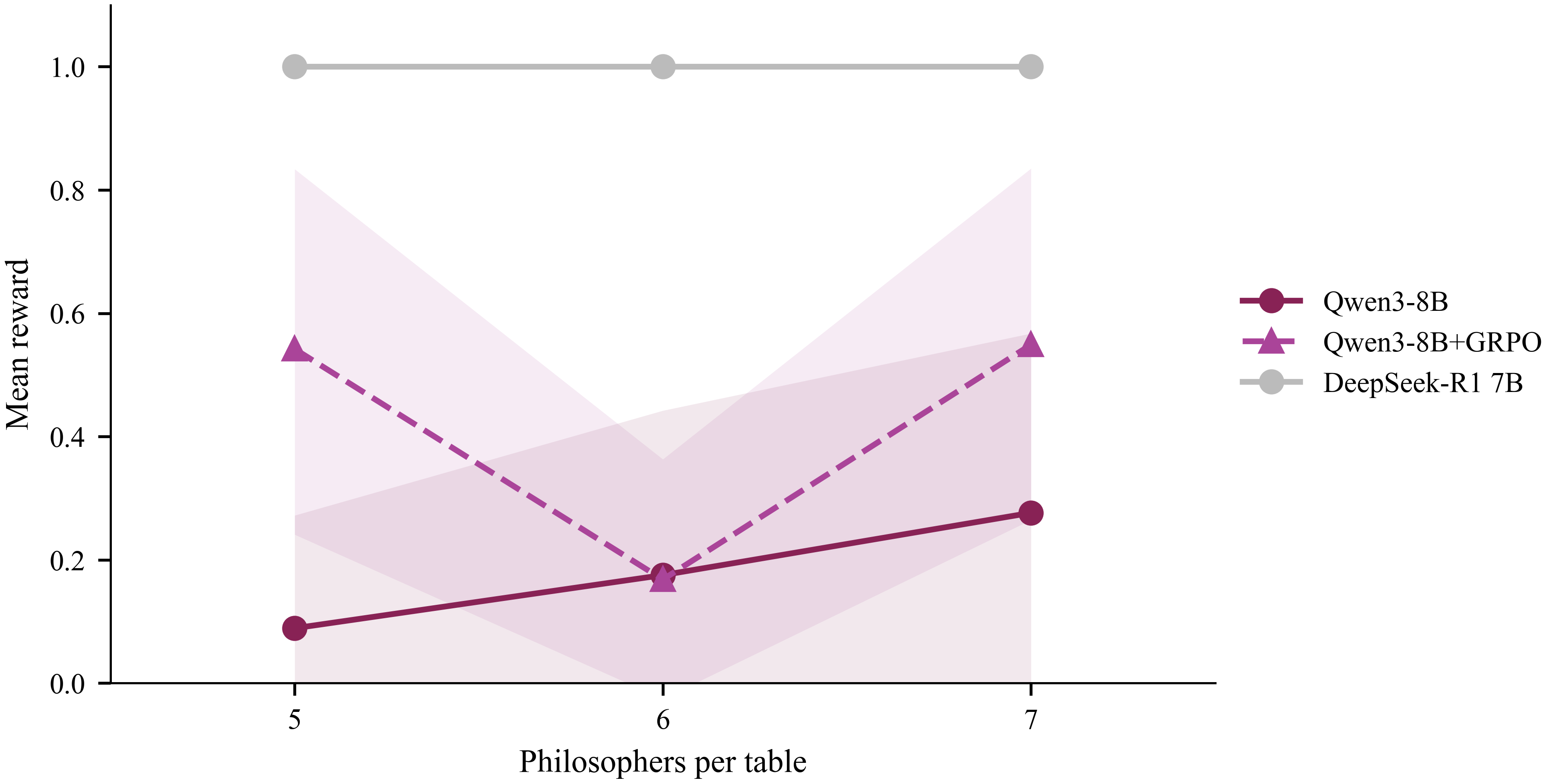}
\caption{Mean reward with five rounds per episode at three philosopher counts. The saved Qwen3-8B+GRPO policy has higher reward point estimates than its base at five and seven philosophers, but a slightly lower estimate at six. DeepSeek-R1-Distill-Qwen-7B receives mean reward 1.0 under the reported parser limitation.}
\label{fig:qwen8b-phil}
\end{figure}

\clearpage
\section{Statistical tests}
\label{app:stats}

Table~\ref{tab:stats} reports Welch's $t$-tests on per-episode reward for the saved Qwen3-14B+GRPO policy versus its base at each philosopher count, together with the small-sample-corrected Hedges' $g$ effect size. None of the three comparisons detects a statistically significant difference at $\alpha = 0.05$, and all reported effect sizes satisfy $|g| < 0.22$. These tests concern mean reward for the evaluated policies. They do not establish equivalence, test throughput or deadlock differences, or estimate variation across independently trained policies.

\begin{table}[H]
\caption{Welch's $t$-test for Qwen3-14B+GRPO versus Qwen3-14B base at each philosopher count. Each row uses 30 episodes per arm. No comparison rejects the null hypothesis at $\alpha = 0.05$.}
\label{tab:stats}
\centering
\begin{tabular}{cccccc}
\toprule
$N$ philosophers & mean (trained) & mean (base) & $t$ & $p$ & Hedges' $g$ \\
\midrule
5  & 0.0924 & 0.1284 & $-0.4432$ & 0.6593 & $-0.1129$ \\
10 & 0.2547 & 0.3500 & $-0.8191$ & 0.4161 & $-0.2088$ \\
15 & 0.2867 & 0.3204 & $-0.2881$ & 0.7743 & $-0.0734$ \\
\bottomrule
\end{tabular}
\end{table}

\clearpage

\end{document}